\documentstyle[aps,preprint]{revtex}
\begin{document}
\title{Correlation  of 
eigenstates in the critical regime of 
quantum Hall systems} 
\author{Krystian Pracz, Martin Janssen, Peter Freche}
\address{ Institut f\"ur Theoretische Physik, Universit\"at zu
K\"oln, Z\"ulpicher Strasse 77, 50937 K\"oln, Germany}  
\date{28.04.96}
\maketitle
\begin{abstract}
We extend the  multifractal analysis
of the statistics of critical 
wave functions in quantum Hall systems by calculating numerically the 
 correlations  of local 
amplitudes corresponding to eigenstates
at two  different energies.
Our results confirm multifractal scaling
relations which are different from
those occurring in conventional critical phenomena.
The critical exponent corresponding to the typical
amplitude,
 $\alpha_0\approx 2.28$,
gives an almost complete characterization of the critical behavior
of eigenstates, including correlations.
 Our results support the interpretation of the local density of states
being an order parameter of the Anderson transition.
\end{abstract}
\section{Introduction}\label{secint}
Two-dimensional independent electrons in the presence of static disorder
and a strong magnetic field undergo  disorder induced
localization-delocalization (LD) transitions when the Fermi energy crosses
critical energies -- the  Landau energies.
 These LD transitions are believed to be responsible
for the  integer quantum Hall effect 
\cite{PraGirv,JanB}.
In finite size systems the localization length $\xi$ 
of the electronic states
is larger than the system size $L$ for a certain energy  range, $\Delta
E$,
 around the
critical energies. These states are called {\it critical states}.  In the
thermodynamic limit 
$\Delta E\propto L^{-1/\nu}$ where $\nu$ is the critical exponent of $\xi$.
An obvious task for a theory  of the LD transition  is to yield   
the statistics and scaling behavior of eigenstates in the critical regime.
After  pioneering works by Wegner \cite{Weg80}
and  Aoki \cite{Aok83Aok86} it became clear
that the critical wave functions have a  multifractal structure
(for a review see \cite{JanR} and references therein). 
The entire distribution of local amplitudes and its scaling behavior
is encoded in the multifractal $f(\alpha)$ spectrum. The distribution
is broad on all length scales and close to a log-normal distribution.
The most important 
quantity is the maximum position, $\alpha_0$, of $f(\alpha)$.
It
 describes the scaling behavior of the geometric mean 
of what serves as a {\it typical} amplitude of a critical wave function.
 The $f(\alpha)$ spectrum has
been
interpreted as a spectrum of critical exponents related to the
order
parameter field of the LD transition \cite{JanR}.  To be consistent
with such an interpretation $f(\alpha)$ has to share several features
with  scaling exponents in conventional critical phenomena. Firstly, it has to
be universal, i.e. independent of the disorder configuration and the
microscopic details of the electrons state. In previous studies this  was
confirmed \cite{Pok91Huc92Kle95}.
Secondly, correlations of the order parameter field have to be related
to $f(\alpha)$ by appropriate scaling relations. A systematic
investigation
of this topic is presented here (Related  conformal
scaling relations were  recently investigated by Dohmen et al.
\cite{Doh95}). Our numerical data are consistent with universal
 scaling relations between
$f(\alpha)$
and
 scaling exponents of the energetic and spatial correlations
of critical wave functions.

\medskip
The article is organized as follows.
 In  Sec.~\ref{secdis} we explicitly demonstrate
that the multifractal
spectrum describes the distribution function of local amplitudes.
In Sec.~\ref{secfix}  the spatial correlations of 
local amplitudes (and powers thereof) of
 wave functions at a fixed energy in the critical regime are investigated.
We study the scaling with respect to the size of boxes which the
local amplitude is averaged over, and with respect to the distance 
between correlated amplitudes.  We find that scaling relations
are fulfilled which can also be obtained by heuristic arguments.
In Sec.~\ref{sectwo} 
 we study the correlator of two local amplitudes corresponding to
critical eigenstates at  energies separated by $\omega$.
We find that a characteristic length scale
$L_\omega$ (introduced by Chalker \cite{ChaD,ChaPA})
serves  as a cutoff length for multifractal
correlations. Remarkably,
 two local amplitudes corresponding to eigenstates
separated in energy by $\omega$  are correlated in
the same way as two amplitudes corresponding to one eigenstate,
provided their spatial distance  is less than $L_\omega$.
 By transforming from the wave function amplitudes to
the local density of states our results show that 
 the local density of states has several features in common with
order parameter fields.  
The conclusions are summarized in Sec.~\ref{seccon}.

\section{Distribution of local amplitudes}\label{secdis}
In this section we first review the multifractal analysis
of the distribution of local amplitudes of critical eigenstates.
Secondly, we confirm that the distribution is encoded in the
multifractal $f(\alpha)$ spectrum  by a direct comparison of the
numerically obtained histogram of local amplitudes with the
theoretical
distribution function that follows from the numerically obtained
$f(\alpha)$
spectrum.  

\medskip

In 1983 Aoki \cite{Aok83Aok86}
 gave a nice argument for the multifractal behavior of
critical wave functions (although at that time the phrase
`multifractality'
was not yet common). His argument goes as follows.
Consider  the inverse participation
number defined by 
\begin{equation}
	{\cal{P}} = \int\limits_{\Omega} {d}^d r\,
	|\psi ({\bf r})|^4  \label{1}
\end{equation}
where $\Omega$ denotes a $d$-dimensional
region with linear size $L$.
If the wave function $\psi ({\bf r})$ is uniformly
distributed -- like in a metallic phase -- then ${\cal{P}}\propto
L^{-d}$ and
the participation ratio $p=({\cal{P}}L^d)^{-1}$ is constant.
In the localized regime ${\cal{P}}\approx \xi^{-d}$ and $p$ vanishes
in the thermodynamic limit.
At the transition point where the wave function is extended
the participation ratio still has to
vanish in the thermodynamic limit if the LD phenomenon is similar
to a second order
phase
transition.
Consequently, ${\cal{P}}$ scales with a power $d^{*}<d$.
Wegner had already  calculated the whole spectrum of exponents
for generalized inverse participation numbers within the non-linear
sigma-model \cite{Weg80}. This spectrum  was interpreted as a multifractal
spectrum by Castellani and Peliti  \cite{Cas86}.
After extensive numerical work the following description of the
statistics
of critical wave functions $\psi(\bf r)$ is now established:

Consider the  box-probability
\begin{equation}
	P(L_b):=\int_{\rm box}|\psi(\bf r)|^2
\end{equation}  
of some box with linear size $L_b$,
  normalized to the total volume $L^d$, $P(L)=1$.
At the LD transition the corresponding distribution function
$\pi(P;L_b/L)$  gives rise to power law scaling for the
moments,  
\begin{equation}
	\left\langle \lbrack P(L_b)\rbrack^q \right\rangle_L 
\propto (L_b/L)^{d + \tau(q)} \, ,\label{2}
\end{equation}
where $d+\tau(q)$ is a non-linear function of $q$.
This non-linearity is a direct
consequence
of Aoki's observation that $d+\tau(2)= d+d^{*}\not=d+d$.
The brackets $\left\langle ...\right\rangle_L$ in Eq.~(\ref{2})
 denote the average over disorder
configurations. In practice it turns out that, to a good accuracy
this average can often  be
substituted by the spatial average over one wave function for a given
configuration. Within  numerical
 accuracy the resulting spectra  are identical.
  This ``universality'' is expected to be
precise in the thermodynamic limit (see also \cite{Mud95}).
  
 The corresponding (universal) distribution function 
can be described in terms of a
single-humped, positive function $f(\alpha)$  called the multifractal
spectrum of the wave 
function,
\begin{equation}
	\pi(P;L_b/L)\, dP \propto (L_b/L)^{d-f(\alpha)} \, d\alpha \, ,
 \label{3}
\end{equation}
where $\alpha:=\ln P/\ln (L_b/L)$;
  $f(\alpha)$ is   related to   $\tau(q)$ by a Legendre transformation
\begin{equation}
	f(\alpha(q))=\alpha(q) q -\tau(q) \; ,\;\;
	\alpha(q)=d\tau(q)/dq\, .\label{4}
\end{equation}
Thus,  the statistics of critical eigenstates
is encoded in $f(\alpha)$ or equi\-valently in $\tau(q)$.
Because of the horse-shoe shape of $f(\alpha)$
it can be approximated by a parabola,
\begin{equation}
	f(\alpha)\approx
	d -(\alpha-\alpha_0)^2/(4(\alpha_0-d))\,.\label{5}
\end{equation}
This parabolic approximation (PA) contains  $\alpha_0$ as the only
parameter besides $d$.  This is due to the assumed validity of
the PA  at least up to $|q|\leq 1$.
Equation (\ref{5}) 
   corresponds to a log-normal
distribution centered around the typical value $P_{\rm typ}:=\exp <\ln P>
 \propto
(L_b/L)^{\alpha_0}$ with log-variance proportional to $\alpha_0-d$.
A simple one-parameter approximation for
$f(\alpha)$ 
 which takes into account that the support
$[\alpha(\infty),\alpha(-\infty)]$
of $f(\alpha)$ is finite, is the semi-elliptic
approximation 
(SEA)
\begin{equation}
	f(\alpha)\approx d\sqrt{1-\frac{(\alpha -\alpha_0)^2}{\alpha_0^2-d^2}}
\, .
\label{6}
\end{equation}

To demonstrate that the  distribution of local amplitudes of
critical eigenstates is encoded in $f(\alpha)$ we present 
 numerical results
for a quantum Hall system (QHS).

\medskip

The 
wave functions are calculated for the 
 model of independent (spinless) electrons
subject to strong magnetic field and  disorder. The disorder was
implemented by a set of $\delta$-impurities with random positions and
random strengths   symmetric around zero.
To avoid that some of the (degenerate)
 wave functions of the pure system have zeroes   precisely at the 
position of the point scatterers  (and thus
are not affected by disorder)
the number of point scatterers was taken to be larger then the number 
of flux-quanta
penetrating the  system's area $L^2$ (cf.~\cite{And}).
The microscopic length scale of the problem is the magnetic length $l_B$
 defined by  the size of a cell
penetrated by a single flux-quantum, $2\pi l_B^2$.
We worked out the representing Hamiltonian matrix in the Landau representation
which is convenient for periodic boundary conditions in one direction
(say $y-$direction). To account for a finite system size in $x-$direction we
adopted the Landau counting procedure which results in a matrix of dimension
$N=L^2/(2\pi l_B^2)$, after projecting to the lowest Landau band.
 The matrix has band structure with a bandwidth of
order $\sqrt{N}$ and allows for diagonalizing systems of linear size
$L$ about  $200 l_B$ with the aid of usual workstations. The
diagonalization yields the eigenvalues and eigenstates for any
desired energy window within the lowest Landau band.
 The determination
of the range $\Delta E$ of critical states was based on a previous
analysis of Thouless numbers within the same model \cite{Fas}.

\medskip

In Fig.~\ref{fig1} the squared amplitudes of a wave function from the center
of the Landau band are shown together with
the $f(\alpha)$ spectrum calculated from these amplitudes.
 The corresponding histogram of
the logarithm of amplitudes (measured on a box of size $4l_B^2$)
is displayed in Fig.~\ref{fig2} together
with the distribution function calculated from the $f(\alpha)$ spectrum
using Eq.~(\ref{3}).  These figures demonstrate that the distribution
of amplitudes is (i) encoded in the $f(\alpha)$ spectrum and (ii)
is close to a log-normal distribution characterized by one critical
exponent
$\alpha_0=2.28\pm 0.03$ (the average over 130 critical states).
Similar findings were presented already in Sec.~12.2 of Ref.~\cite{JanB}.
\section{Correlations at fixed energy}\label{secfix}
In this section we first outline a theoretical
description of scaling relations between  the critical exponents
of the correlations at a fixed energy and the $f(\alpha)$ spectrum.
We follow mainly the presentation of Ref.~\cite{JanR}.
Also, we  present our numerical data which confirm
the scaling relations.

\medskip

To  study the
spatial correlations of amplitudes for a fixed energy 
we consider the $q-$dependent correlations
\begin{equation}
	M^{[q]}(r,L_b,L):=\left\langle \lbrack P_i(L_b)\rbrack^q
\lbrack P_{i+s}(L_b)\rbrack^q\right\rangle_L\label{7}
\end{equation}
where the average is to be taken over all pairs of boxes with fixed distance
$r=sL_b$.

For critical states where the microscopic scale ($l_B$ in our case)
 and the macroscopic scale (the localization
length $\xi$ in our case)  are separated, one can expect power law
behavior of $M^{[q]}$ 
in the regime 
\begin{equation}
	l_B\ll L_b,r,L \ll \xi \, .\label{8}
\end{equation} 
Usually, in critical phenomena one studies correlations for infinite
system size (and $L_b$ being microscopic)
as a function of $r$ alone. This is justified if, for large enough system 
sizes $L$, the correlation function is independent of $L$ (for simplicity
 we neglect 
 any trivial $L$-dependence due to prefactors in the
definition of the observable $P$).
 However,
this is not true in the multifractal case.
 Multifractality reflects broadness of the
distribution function $\pi(P,L_b/L)$ on all length scales. This is due to the
dependence of the box-probability in each box on a large number of
conditions, simultaneously.
More generally, the local box observable $P_i(L_b)$
depends on a large number of conditions for the entire system of
linear size $L$, simultaneously.
 This behavior was denoted as  ``many parameter
(MP) coherence'' \cite{JanR}.
In the context of the LD transition coherence at zero temperatures
is due to  quantum mechanical phase
coherence of the electrons wave function, and disorder introduces a
huge number of parameters, e.g. the position of point-scatterers.
In the case of MP coherence one has to face the fact that 
$M^{[q]}$  depends non-trivially on $L$, even for $L\to\infty$. 
To distinguish between the multifractal and the ``ordinary''
 situations one can implicitly define a  
 length scale $\hat{L}$ by  the requirement
that $M^{[q]}$ will be  independent of 
 $L$ for $L>{\hat{L}}$. We call $\hat{L}$ the MP coherence
length.
Still, two different cases have
to be distinguished. It may happen that $\hat{L}$ introduces a cutoff for
correlations. For example,
the correlation length $\xi$ is a MP coherence length of this kind.
Alternatively, $\hat{L}$ does not introduce a 
cutoff and correlations still show a modified
power law behavior with respect to
$r$ for $r\gg {\hat{L}}$.
Such kind of MP coherence length occurs in
 ordinary critical behavior where $\hat{L}$ is microscopic
and no multifractality occurs, i.e.~$\tau(q)=d(q-1)$.

After these general considerations we can formulate our expectations
for the present case of correlations of critical eigenstates at fixed
energy.
Because of the multifractal character of the critical states whose
range is only limited by the localization length $\xi\gg L$ we
can expect to be dealing with  a  MP coherent situation with
$L$ being the MP coherence length $\hat{L}$, setting the cutoff for
correlations.
Therefore,  we consider the regime $l_B\ll L_b < r < L \ll \xi$ 
and make the ansatz 
\begin{equation}
	M^{[q]}(r,L_b,L)\propto L_b^{x_2(q)}L^{-y_2(q)}r^{-z(q)}\, .\label{10}
\end{equation}

The task is now to find the scaling relations between the
 set of exponents $x_2(q),y_2(q),z(q)$ and  the $\tau(q)$
function of Eq.~(\ref{2}).
These scaling relations can be derived on the basis of heuristic
arguments. Consider the limiting situations (i) where $r$ is of the
order of
 $L_b$ and (ii) where $r$ is of the order of $L$. 
In case (i) the function $M^{[q]}$ will behave like $\left\langle
\lbrack P(L_b)\rbrack^{2q}\right\rangle_L$
while in case (ii) a decoupled behavior occurs $M^{[q]}\sim
\left(\left\langle \lbrack P(L_b)\rbrack^q\right\rangle_L\right)^2$. 
The uniqueness of scaling
exponents allows then to conclude the desired scaling relations
\begin{eqnarray}	
	y_2(q) &=& d+\tau(2q) \label{11a}\\
	x_2(q) &=& 2d + 2\tau(q) \label{11b}\\
	z(q)&=& d +2\tau(q) - \tau(2q)\, \label{11c}
\end{eqnarray}
 already proposed in \cite{Cat,JanR}. 
It is worth mentioning that the sum $x_2(q)-y_2(q)-z(q)$
vanishes due to the normalization of the wave function.
To verify numerically the scaling relations it is thus 
enough to verify two of the three equations given above.
Therefore, we concentrate on the exponents $z(q)$ and $x_2(q)$ for
which
we can set a fixed system size in  numerical calculations.
This
reduces the computational effort substantially.

Let us summarize the analytic behavior of   $x_2(q)$ and $z(q)$
according to Eqs.~(\ref{11b}, \ref{11c}):
$x_2(q)$ is a monotonic increasing function with negative curvature
and asymptotic slopes given by $2\alpha (\mp \infty)$.
It vanishes at $q=0$.  $z(q)$ is non-negative with minimum at $(0,0)$.
For $q>0 (<0)$ it is monotonically increasing (decreasing) and is
asymptotically bounded by the dimension $d$ (see Fig.~\ref{fig3}).
To check on the validity of Eq.~(\ref{11c})
we took $100$ critical states of a system with $L=200 l_B$
and calculated $M^{[q]}(L_b,r,L)$ with fixed values $L_b=l_B, 4 l_B$;
$L=200l_B$. The distance $r$ was varied 
from $L_b$ to $150 l_B$.  The periodic boundary conditions in 
$y$-direction reduce the upper scale for
a reliable determination of exponents to   $r <\approx 100 l_B$.
As can be seen from Fig.~\ref{fig4} the power law behavior holds up to
$\approx 60 l_B$. The numerical data for $z(q)$ were obtained by
 determining the
slope in the linear regime of the plots $\ln M^{[q]}$ vs. $\ln r$.
In Fig.~\ref{fig5} the average of
$z(q)$ data
over $100$ states  is shown in the regime $|q|<2$. For comparison the 
function $d+2\tau(q)-\tau(2q)$ is plotted, too. 
 Within the errors  the validity of the
 scaling relation Eq.~(\ref{11c})  can be confirmed.
For later comparison we mention that $z(1)=0.43\pm 0.05$, $z(0.5)=0.13\pm 0.03$
and $d^2z/dq^2 (q=0)=1.1 \pm 0.08 \approx 4(\alpha_0-d)$. 
To determine the exponents $x_2(q)$ numerically 
we fixed $r= 100 l_B$, $L=200l_B$ and varied $L_b$ from $2l_B$ to $r$.
As is shown in Fig.~\ref{fig6} the scaling relation Eq.~(\ref{11b})
is fulfilled, too.

In summary, we have presented numerical evidence for the validity of
the scaling relations Eqs.~(\ref{11b},\ref{11c}).
\section{Correlations at two different energies}\label{sectwo}
In this section we firstly discuss the scaling relations for
correlations
of critical eigenstates at two different energies. We stress the role
of a length scale $L_\omega$ related to the energy separation
$\omega$.
Secondly, we present numerical calculations which
lead to the  interpretation of $L_\omega$ being
a cutoff scale for MP coherence. Thirdly, we verify the
validity
of scaling relations based on this interpretation. Finally, we discuss
the relevance of our results  for the interpretation of the local
density of states as being an order parameter of the LD transition.

\medskip

We define the $q$-dependent correlation of box probabilities corresponding to 
two different eigenstates with energies $E$ and $E+\omega$
\begin{equation}
	M_{\omega}^{[q]}(r,L_b,L):=\left\langle \lbrack P_i(E;L_b)\rbrack^q
	\lbrack P_{i+s}(E+\omega;L_b)\rbrack^q\right\rangle_L \, .\label{13} 
\end{equation}
To understand the correlation behavior of non-localized
states with respect to the
energy separation one has to
compare the relevant energy scales of the problem.
These are the average level spacing $\Delta$ and the 
energy $E_c(\omega)$  corresponding to the time a wave packet
(formed from states within an energy window of width $\omega$) needs to diffuse
through the system, $L^2=(\hbar/E_c(\omega))D(\omega)$.
Here $D(\omega)$ is the corresponding diffusion constant.
 According to Chalker \cite{ChaPA},
 these scales give rise to the definition of two
length scales depending on  the energy separation $\omega$:
\begin{eqnarray}
	{\tilde{L}}_\omega & := & (\omega/E_c(\omega))^{-1/2}L \label{14}\\
	 L_\omega & := &
(\omega/\Delta)^{-1/d}L \, .\label{15}
\end{eqnarray}
The first length scale, $\tilde{L}_\omega$,
is the typical distance a wave packet will travel diffusively in a time
$\hbar/\omega$. From this it is natural to assume
that correlations between $P_{i}(E;L_b)$ and $P_{i+s}(E+\omega;L_b)$
will be present at least for distances $r \ll {\tilde{L}}_\omega$ whereas for
 larger distances the amplitudes are  uncorrelated.
Such uncorrelated behavior  is typical for
 random matrix theory approaches to
 chaotic systems. The corresponding assumption is known
as
`isotropy' or `no-preferential basis' assumption and  means that
the unitary matrices that diagonalize the Hamiltonian are distributed
uniformly in the unitary group and no correlations (apart from the unitarity
property) between different
matrix elements occur. Thus, the
presence of correlations  means a  breakdown of  the `no-preferential
basis' 
assumption. In electron systems with spatial
disorder, however, a preference
to some basis is always given. This preference
 is lost in $M^{[q]}_\omega$ for distances $r\gg \tilde{L}_\omega$.

The second length scale,  $L_\omega$, is the linear  size of a system
with
level spacing $\omega$. Taking the existence of a preferential
basis for being significant  
we will adopt the hypothesis that two
  wave functions with energetic separation smaller
than
 the level spacing show a spatial correlation behavior
of its amplitudes similar to
that corresponding to  one of those wave functions, i.e.
they are statistically
indistinguishable. 

 At the critical
point of the LD transition the conductance becomes a size independent
quantity, $G=g^{*}e^2/h$,   and with the help of the Einstein relation 
between conductivity and diffusion one finds 
${\tilde L}_\omega= (g^{*})^{1/d}L_\omega$ \cite{ChaPA}.
Since $g^{*}$ is of ${\cal O}(1)$ the two length scales coincide at the LD
transition. Therefore, we will  focus our considerations
on the role of $L_\omega$ and start with the following
working hypothesis:

\smallskip

{\it $M^{[q]}_\omega$ will show the same correlation exponent $z(q)$
as $M^{[q]}_{\omega=0}$ provided  $r\ll L_\omega$.
For distances $r\gg L_\omega$ the amplitudes  $P_{i}(E;L_b)$ and
$P_{i+s}(E+\omega;L_b)$ are uncorrelated.}

\smallskip

\noindent
This means that $L_\omega$ is a MP phase coherence length
setting a cutoff for correlations.
Asking for the scaling properties of $M^{[q]}_\omega$
in the regime $L_b < r < L_\omega \leq L$
we make the  ansatz (cf.~Eq.~(\ref{10}))
\begin{equation}
	M^{[q]}_\omega\propto
L_b^{X_2(q)}r^{-z(q)}L_\omega^{Z(q)}L^{-Y_2(q)}\, .\label{16} 
\end{equation}
Here we have already anticipated that the exponent with respect to $r$
is  $z(q)$, as given before and will check if this is consistent
within the following procedure of deriving the scaling relations
between
$X_2(q)$, $z(q)$, $Z(q)$, $Y_2(q)$ and $\tau(q)$.
As in the case of $M^{[q]}$ for a fixed energy we consider limiting
situations.
In case (i) we keep $r$ of the order of $L_b$ and $L_\omega$ of the order
of $L$, whereas 
in case (ii) we keep $r$ of the order of $L_\omega$.
 Following our working hypothesis  for  case (i)
we expect no difference to the case with zero energy separation 
\begin{equation}
	M^{[q]}_\omega \propto \left\langle \lbrack P_i(E;L_b)\rbrack^{2q}
	\right\rangle_L \, ,\label{17}
\end{equation}
 resulting in
\begin{equation}
	X_2(q)-z(q)=d+\tau(2q)= Y_2(q)-Z(q) \, .\label{18}
\end{equation}
For case (ii), according to the hypothesis, $M^{[q]}_\omega$
 depends neither on $r$ nor on $L_\omega$ and  
approaches the uncorrelated
value,
\begin{equation}
      M^{[q]}_\omega \propto \left(\left\langle \lbrack P_i(E;L_b)\rbrack^{q}
	\right\rangle_L\right)^2\, .
\label{19}
\end{equation}
This leads to 
\begin{equation}
     z(q)=Z(q) \; ; \;\; X_2(q)=2d+2\tau(q)=Y_2(q) \, .\label{20}
\end{equation}
Equations (\ref{18},\ref{20}) yield the scaling relations
\begin{eqnarray}
       X_2(q)=2d+2\tau(q)=Y_2(q) \label{21}\\
       z(q)=Z(q)=d+2\tau(q)-\tau(2q) \label{22}
\end{eqnarray}
which form a central result of this article. Now the following conclusions can
be drawn:
1.  The result
for $z(q)$ is the same as in the case of  zero energy separation.
2. The energy separation $\omega$ is not an independent scaling parameter
 but appears only in the combination $L_\omega/r$. 
3.  The exponent
corresponding to the box size, $x_2(q)=X_2(q)$, is not affected by a finite
energy separation but the exponent
corresponding to the system size $L$ (which is $y_2(q)$ for zero energy
separation) 
 splits up into the exponents $Z(q)$ (corresponding to $L_\omega$)
and   $Y_2(q)$ (corresponding to $L$ for finite energy separation).

 In Ref.~\cite{JanR} it has been 
speculated that for $r\gg L_\omega$ modified
power law correlations may still exist.
 Such
behavior
would be in conflict with the interpretation of ${\tilde L}_\omega$
adopted here  and it would  lead to different
scaling relations in comparison to Eqs.~(\ref{21},\ref{22}). Especially, the
equality between $z(q)$ and $Z(q)$ would be lost for arbitrary values
of $q$. However, the possibility of such behavior   could not be ruled
 out on the basis of previously obtained
 numerical data which correspond to the case
$q=1$.

We will now demonstrate that the working hypothesis formulated above and 
the resulting conclusions 1. and 2. are  
consistent with our data.

\medskip

Firstly, we checked  if $L_\omega$ serves as a cutoff
for correlations. That this is indeed the case can be seen from 
Fig.~\ref{fig7}  showing the logarithm of $M^{[q]}_\omega$
as a function of $\ln r$ ($r$ is measured in units of $l_B$)
for a fixed size $L_b=l_B$ and  $q=0.5$.
The energy separation $\omega$ corresponds to 
$L_\omega = 30 l_B$. A  clear $r$-dependence up to approximately this scale
can be observed.  In order to calculate a reliable scaling exponent
$z(q)$ one has to take $L_\omega \geq 100 l_B$. Doing so, we found
(within the errors) 
the same $z(q)$ values as for correlations with no energy separation,
e.g. $z(1)=0.39\pm 0.04$, $z(q=0.5)=0.10\pm 0.03$. 
 Thus, the working hypothesis turned out
to be consistent with our data.   
The next step is  to investigate the scaling
exponent $Z(q)$ corresponding to $L_\omega$ 
which should, according to the scaling relation Eq.~(\ref{22}), be
equal to $z(q)$, resulting in  a combined scaling parameter $r/L_\omega$.

For the numerical determination of $Z(q)$ we used about $130^2$ combinations
of pairings between different eigenstates.  For each pair the actual value
of the cutoff scale for correlations fluctuates around the calculated
value $L_\omega$. This fact requires a large amount of data to
extract reliable scaling exponents. From the plots of $\ln M^{[q]}_\omega$
versus $\ln L_\omega$
with fixed values of $r\approx L_b=l_B$ we determined the
approximate linear behavior in a 
regime between $L_\omega\approx 20 l_B$ and $L_\omega \approx 50 l_B$.
As can be  seen in Fig.~\ref{fig7} ($q=0.5$) there are 
fluctuations  due to rare pairings.
In Fig.~\ref{fig8} we show the numerically obtained $Z(q)$ function
together with  the errors  of the linear fit in log-log plots.
It is compared with the previously obtained $z(q)$ function for
zero energy separation (cf.~Sec.~\ref{secfix}).
The data show that the equality $Z(q)=z(q)$ is consistent with our
data, e.g.~$Z(1)=0.38\pm 0.04$, $Z(0.5)=0.09\pm 0.03$ and 
$d^2 Z/dq^2 (q=0)=1.0\pm 0.1$.

We have thus demonstrated that the working hypothesis as well as the
scaling relations are consistent with our numerical results.

\medskip

Having established the role of $r/L_\omega$ as the relevant scaling parameter
for correlations of eigenstate amplitudes 
with universal exponent $z(q)$ (related  to $\tau(q)$ by a scaling relation)
let us now discuss the consequences of these findings for the interpretation
of the local density of states being an order parameter of the LD transition.
The local density of states (LDOS) is defined  formally as
$
 \rho(E,{\bf r})=\sum_{n} \delta(E-E_n) |\psi_{n}({\bf r})|^2 
$
where $\psi_n({\bf r})$ are wave functions corresponding to eigenenergies $E_n$.
In a finite system this function has isolated peaks at  $E_n$ and
becomes a smooth  function of energy only
for an open system or in the thermodynamic limit.
 This is also true 
for the global
density of states (DOS) $\rho(E)=L^{-d}\sum_n \delta(E-E_n)$.
Since the smallest relevant energy scale for the structure of the DOS as well
as of the LDOS   is set by the average
level spacing $\Delta$ a smearing out of the $\delta$-functions over this scale
is needed to talk about DOS and LDOS in finite systems. It is known that
the average DOS does not reflect the LD transition but is a smooth
function of energy  and independent of
system size $L$. With this smearing out
of the $\delta$-functions we define the 
LDOS  as
\begin{equation}
   \rho(E,{\bf r}) = \Delta(E)^{-1}|\psi(E,{\bf r})|^2 \, \label{23}
\end{equation}
where $|\psi(E,{\bf r})|^2$ stands for the microcanonical average of
squared amplitudes at a given energy $E$.
Since $\Delta(E)$ is smooth
in energy and behaves as $L^{-d}$,  the  scaling behavior of the LDOS
is determined by that of the wave function.
Consequently, we have
\begin{equation}
    \left\langle \lbrack \rho(E,{\bf r})\rbrack^q
	\right\rangle_L \propto L^{(q-1)d -\tau(q)}\, \label{24}
\end{equation}
and for  the typical value
\begin{equation}
\rho_{\rm typ} = \exp [\left\langle  \ln (\rho({\bf r}))\right\rangle_L] \propto L^{d-\alpha_0}
\label{25}
\end{equation}
which {\it does} reflect the LD transition.
 Scaling $L$ with the localization length
$\xi\propto |E-E^{*}|^{-\nu}$ where $E^{*}$ and $\nu$ are the
critical energy and the critical exponent of the localization length,
respectively, we arrive at the conclusion that the typical LDOS
vanishes on approaching the critical energy with exponent 
$\beta_{\rm typ}=\nu(\alpha_0-d)$.
This observation has led to the interpretation of the LDOS being an
order parameter field of the LD transition \cite{JanR} (see also
\cite{Weg80,Mir91}).
The unconventional feature as compared to ordinary critical phenomena
lies in the facts that (i) the order parameter field has a broad distribution
resulting in a non-linear dependence of exponents on the degree of
moments considered (multifractality) and (ii) that the average value
has vanishing scaling exponent  while the typical value  gives rise to
a positive scaling exponent $\beta_{\rm typ}$.
The scaling relations that we derived for the wave functions
amplitudes transform to scaling relations of the LDOS since
each box amplitude has to be multiplied by a constant factor of $L^{d}$,
\begin{eqnarray}
	\left\langle (\rho(E,{\bf r}_1))^q 
	(\rho(E+\omega,{\bf r}_2))^q\right\rangle_L \propto
	(L_\omega/r)^{z(q)} L^{-A(q)} \; ;\;\; r=|{\bf r}_1-{\bf r}_2|
	\label{26}\\
	z(q)=d+2\tau(q)-\tau(2q)\; ; \;\; A(q)=2(1-q)d + 2\tau(q)\, .\label{27}
\end{eqnarray}
In this light our scaling relations turn out to be the appropriate
scaling relations connecting the spatial correlations
of the local order parameter field to its scaling dimensions
(cf. Eqs.~(\ref{24}),(\ref{26}),(\ref{27})). 
However, due to the MP coherence these scaling relations are different
from those of  ordinary critical exponents where only one
correlation exponent appears and no multifractality has to be taken
into account. The most remarkable difference lies here in the
observation that power law scaling is also present in the MP
coherence length $\hat{L}$. The latter is given by the system size $L$
for energy separation  smaller than the level spacing
or by $L_\omega$ for energy separation larger than the level spacing.
Our findings demonstrate that the interpretation of the LDOS being an
order parameter field for the LD transition is supported by the
existence of universal scaling relations. 
We  close this section by pointing out  that
 $z(1)=2-\tau(2)\approx 0.4\not= 0$ 
(for the correlator of the density
of states) with $L_\omega/r$ forming the scaling parameter is equivalent
(cf.~\cite{ChaPA,HucSCH94})
to the phenomenon of  ``anomalous diffusion''  found by
Chalker and Daniell \cite{ChaD}.
 As pointed out in \cite{ChaD},
the anomalous character of diffusion lies
in the non-Gaussian dispersion of a wave packet in time $t$ despite the fact
that the average diameter grows like $\sqrt{t}$. This non-Gaussian
time dispersion is caused by the multifractal character of eigenstates.
\section{Conclusions}\label{seccon}
We have demonstrated explicitly that the distribution of amplitudes
of critical eigenstates in quantum Hall systems (QHS) is contained
in the multifractal $f(\alpha)$ spectrum,  essentially
characterized
by one critical exponent, $\alpha_0\approx 2.28$ (Sec.~\ref{secdis}).
 Following Ref.~\cite{JanR} 
we derived scaling relations which relate the critical exponents 
of the spatial correlation of amplitudes (for a fixed energy in the
critical regime)  
to the multifractal spectrum (Eqs.~(\ref{11a},\ref{11b},\ref{11c}))
and demonstrated that they are consistent with numerical data obtained
for  QHS (Sec.~\ref{secfix}). Most interesting is
the (non-trivial) dependence of the correlator on the system size.
In section~\ref{sectwo} we considered correlations of amplitudes
corresponding to critical eigenstates with energy separation $\omega$.
Following Chalker \cite{ChaPA} we identified a length scale $L_\omega$
(describing the system size with level spacing $\omega$) as the
relevant
cutoff for correlations. Furthermore, we exploited the hypothesis that the
amplitudes are correlated as for zero energy separation provided their
distance is much less than $L_\omega$. We found scaling relations
(Eqs.~(\ref{24},\ref{26},\ref{27})) which relate all  the 
correlation exponents to the universal multifractal spectrum of
critical
 eigenstates. Most important are (i) the confirmation of the 
 scaling parameter $L_\omega/r$ and (ii) the identification of $L_\omega$
as the upper limit for multifractal correlations. We discussed
implications
of our
results for the statistical properties of the local density of states
in the critical regime. 
Our findings demonstrate that the interpretation of the LDOS being an
order parameter field for the LD transition is supported by the
existence of universal scaling relations. 

\bigskip

\centerline{{\bf Acknowledgments}}

We thank A. Altland, J. Hajdu and B. Huckestein for useful
discussions.
This work has been supported by the Sonderforschungsbereich 341 of the
Deutsche Forschungsgemeinschaft.

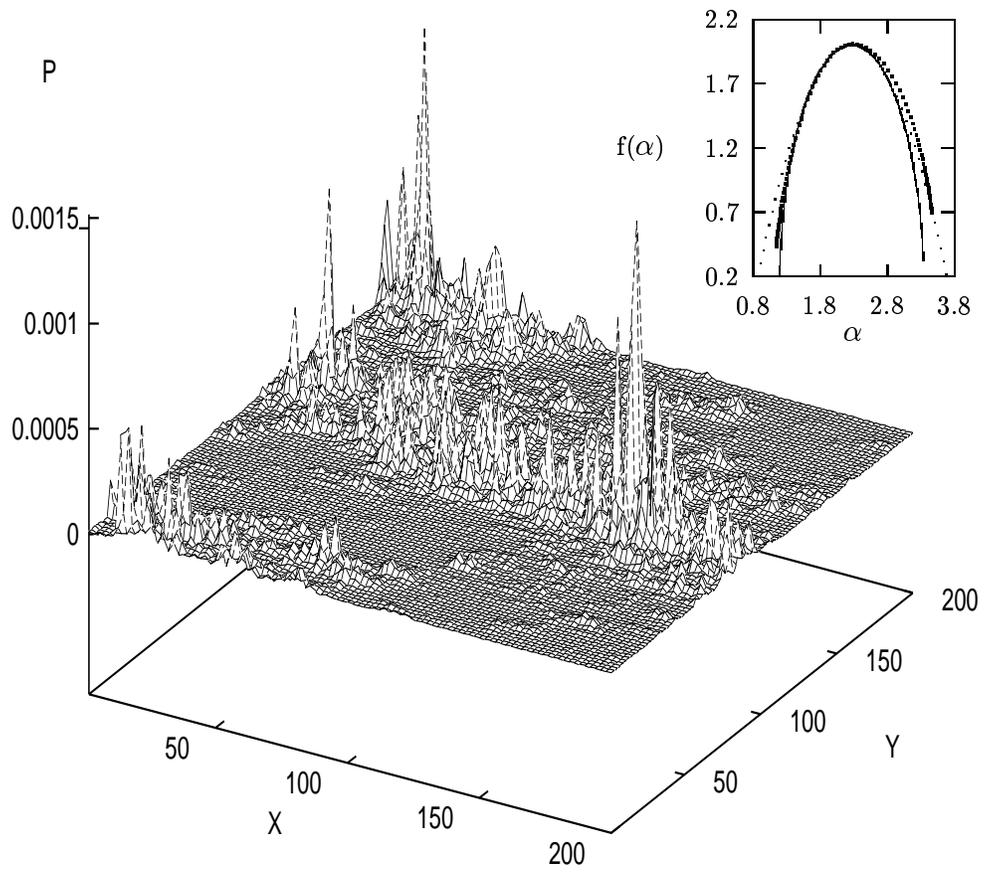
\begin{figure}
\caption{Squared amplitudes of a critical wave function for a system of linear
size \protect{$200$} magnetic lengths. The corresponding 
\protect{$f(\alpha)$} spectrum ($\protect\bullet$)
 together with the parabolic
approximation ($\protect\cdots$)
(PA, \protect{Eq.~(\ref{5})}) and the semi-elliptic approximation  
(--) (SEA,
\protect{Eq.~(\ref{6})}) are also shown.} 
\label{fig1}
\end{figure}
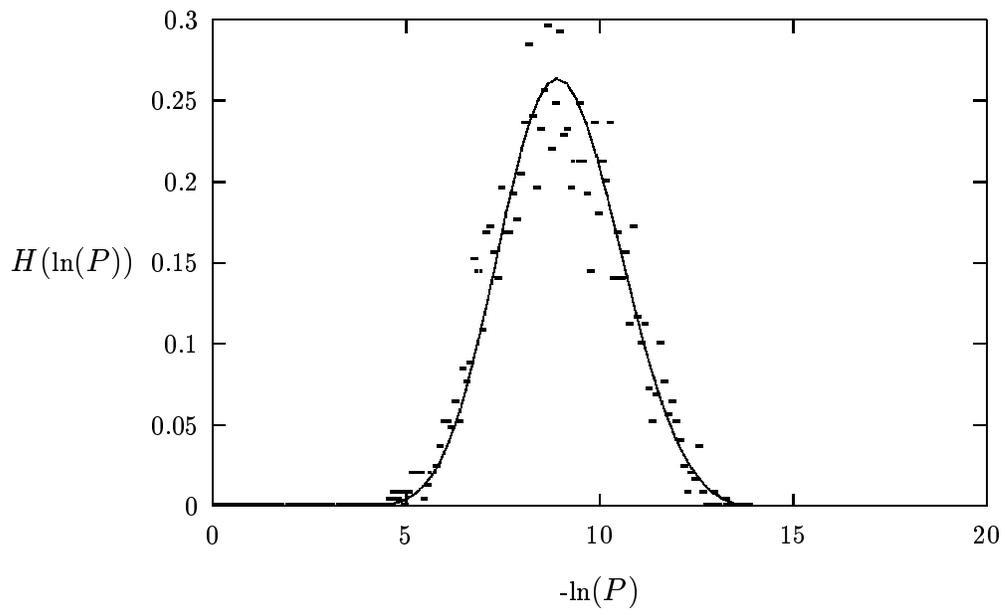
\begin{figure}
\caption{Histogram of the logarithm of squared amplitudes shown in
Fig.~\ref{fig1}. The continuous curve is the distribution function
following from the corresponding $f(\alpha)$ spectrum via Eq.~(\ref{3}).}
\label{fig2}
\end{figure}
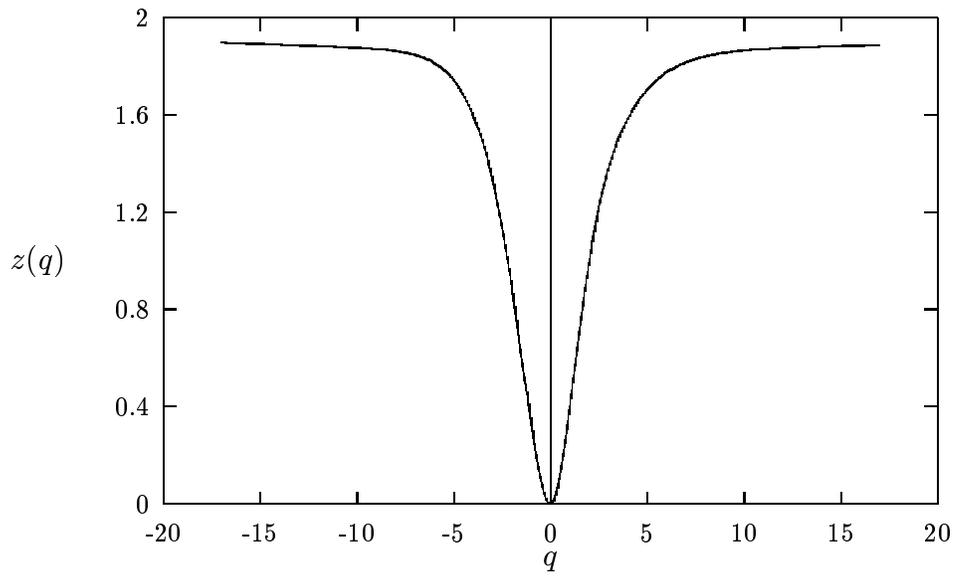
\begin{figure}
\caption{The function of critical exponents $z(q)$ following from the
the scaling relation
Eq.~(\ref{11c}).}
\label{fig3}
\end{figure}
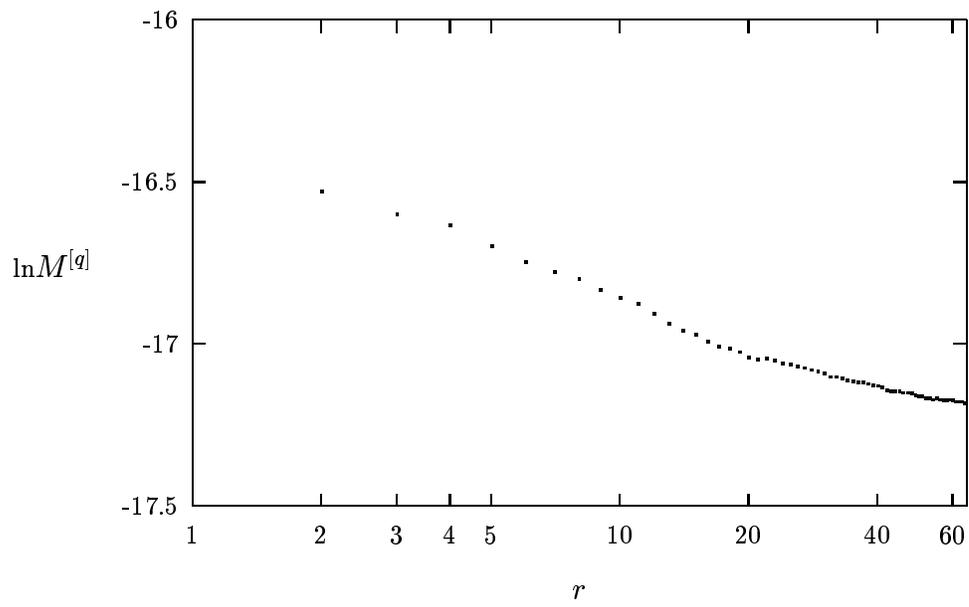
\begin{figure}
\caption{The correlator $M^{[q]}$ (for $q=0.5$) as a function of the spatial
distance $r$ in a log-log plot.}
\label{fig4}
\end{figure}
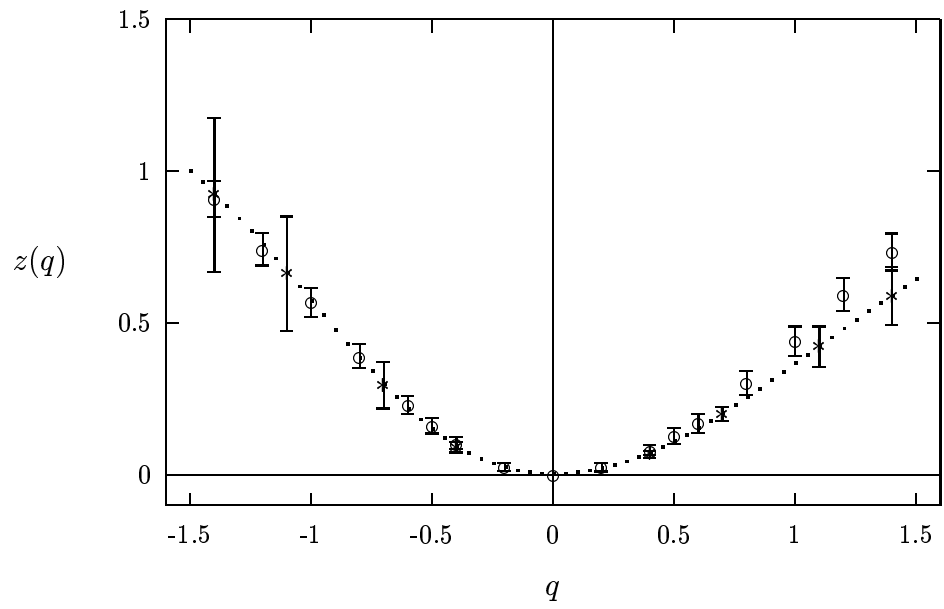
\begin{figure}
\caption{The numerically obtained data for $z(q)$ ($\ast$)
in comparison with data following 
from the scaling relation Eq.~(\ref{11c}) ($\circ$).}
\label{fig5}
\end{figure}
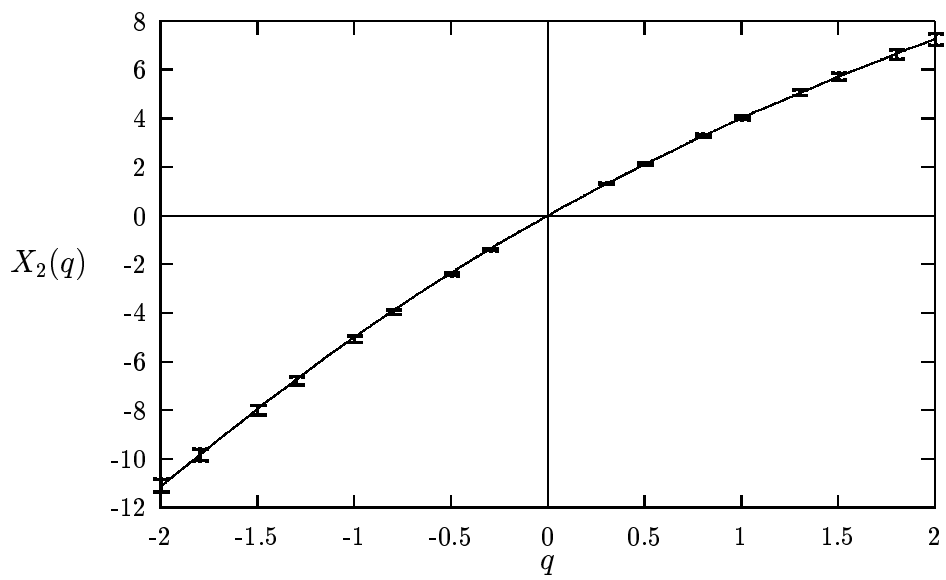
\begin{figure}
\caption{The numerically obtained data for $x_2(q)$. The line shows 
 the data following from the scaling relation 
Eq.~(\ref{11b}).}
\label{fig6}
\end{figure}
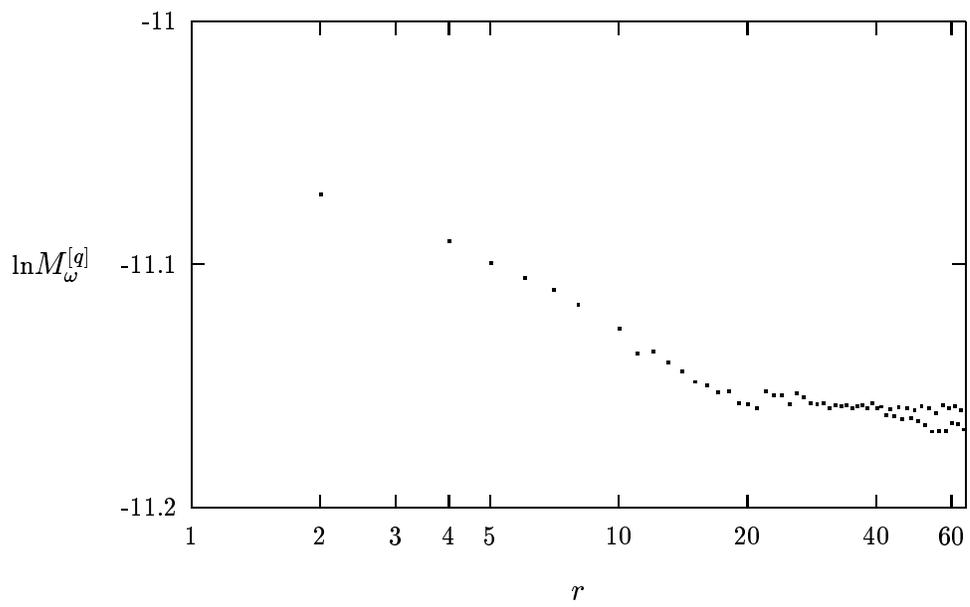
\begin{figure}
\caption{The correlator $M^{[q]}_\omega$ (for $q=0.5$) as a function
of the spatial distance $r$ in a log-log plot for
$L_\omega = 30 l_B$.}
\label{fig7}
\end{figure}
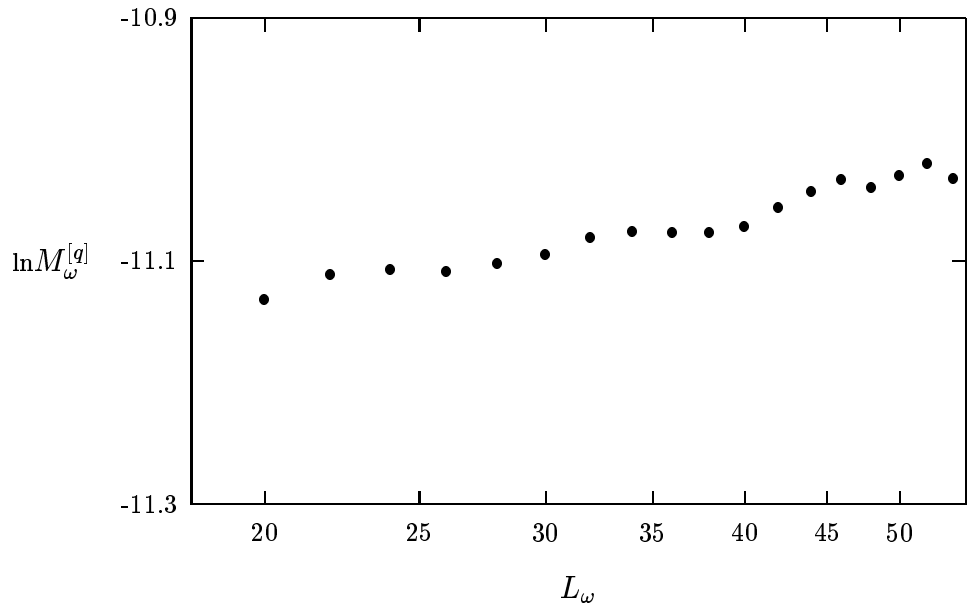
\begin{figure}
\caption{The correlator $M^{[q]}_\omega$ (for $q=0.5$) as a function
of the length scale $L_\omega$ in a log-log plot.}
\label{fig8}
\end{figure}
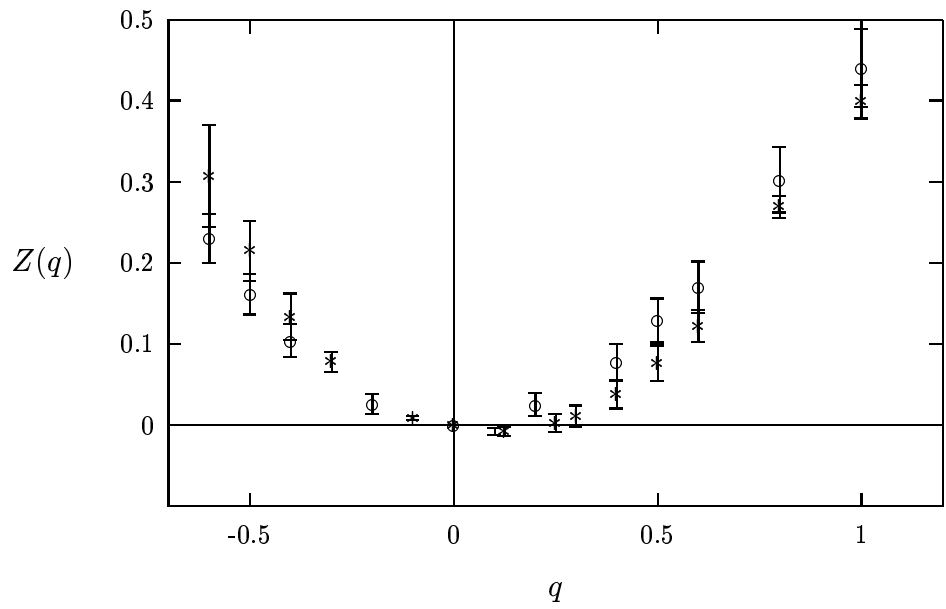
\begin{figure}
\caption{The numerically obtained data for 
$Z(q)$ ($\circ$)
in comparison with the numerically obtained data for $z(q)$ ($\ast$).}
\label{fig9}
\end{figure}

\end{document}